\documentclass[preprint,aps]{revtex4}
\usepackage{color}
\usepackage{mathtools}
\usepackage{diagbox}
\usepackage{appendix}
\usepackage{amsmath,amssymb,amsfonts,dcolumn,color,graphicx,graphics,latexsym,epsfig}
\usepackage[compatibility=false]{caption}
\usepackage{subcaption}
\usepackage{hyperref}
\usepackage{natbib}
\usepackage{float}
\usepackage{booktabs}
\def\beq{\begin{equation}}
\def\eeq{\end{equation}}
\def\barr{\begin{array}}
\def\earr{\end{array}}

\usepackage{caption}
\usepackage{subcaption}

\begin{document}
\title{A curious variant of Bronnikov-Ellis spacetime}

\author{
Sayan Kar}
\email{sayan@phy.iitkgp.ac.in}
\affiliation{Department of Physics, Indian Institute of Technology
Kharagpur, 721 302, India}

\begin{abstract}
\noindent We explore a curious but simple variant of the
Bronnikov-Ellis (BE) wormhole spacetime with
a specific `red-shift function' (i.e. $g_{00}$) in the line element.
The matter required to support such a geometry
violates
the local Null Energy Condition (NEC) only around the throat and the global Averaged Null Energy Condition (ANEC) integral along radial null geodesics may be adjusted to
 arbitrarily  small negative values, using
 metric parameters. Properties of the 
line element manifest in the metric functions, curvature
and the required matter stress energy
are delineated.
Further, exact null and timelike geodesics are found and 
generic features of periodic/non-periodic motion (closed, bounded 
or open) are presented.
Scalar wave propagation 
is also solved analytically, thereby 
providing a partial check on the stability of the geometry under 
scalar perturbations.
Interestingly, we note that this BE variant may be viewed
as a four dimensional, timelike section of a five dimensional, static,
non-vacuum, Witten bubble--like geometry
which, with an extra dimension, also has wormhole features and is threaded by  
matter satisfying the NEC.

\end{abstract}

\maketitle

\section{\bf Introduction} 

\noindent The Bronnikov--Ellis (BE) spacetime 
\cite{bronnikov,ellis}
has been studied extensively over many years, by numerous authors. It is known as an example of a Lorentzian wormhole geometry. Even though
the matter--stress-energy required to support this geometry
consists of an energy--condition violating, negative kinetic energy
scalar field, the simplicity of the line element (at least in a special case) 
is perhaps a reason behind its popularity. It is also true that not many
`exact solutions' are known for Lorentzian wormholes and
BE presents a useful example.

\noindent Let us first recall the line element. It is given as:
\begin{eqnarray}
ds^2 = - e^{f(l)} dt^2 + e^{-f(l)} \left [ dl^2 + \left (b_0^2 + l^2\right ) d\Omega_2^2\right ]
\end{eqnarray}
where $f (l) = \frac{M}{b_0} \left ( \arctan \frac{l}{b_0} -\frac{\pi}{2}\right )$.

\noindent The phantom (negative kinetic energy) scalar field configuration $\phi (l)$ is:
\begin{eqnarray}
\phi (l) = \frac{D}{b_0} \left ( \arctan \frac{l}{b_0} -\frac{\pi}{2}\right )
\end{eqnarray}
where, we have $4 D^2 = M^2 + 4 b_0^2$.

\noindent A simple special case arises for $M=0$, for which  we have $D= b_0$ and the line element becomes:
\begin{eqnarray}
ds^2 = -dt^2 + dl^2 + \left ( b_0^2 + l^2\right ) d\Omega_2^2
\end{eqnarray}
Using the standard `$r$' coordinate, we can rewrite the above line element as
\begin{eqnarray}
ds^2 = -dt^2 + \frac{dr^2}{1-\frac{b_0^2}{r^2}} +  r^2 d\Omega_2^2
\end{eqnarray}
The $t$=constant, $\theta=\frac{\pi}{2}$ sections of the $M=0$ BE spacetime are catenoids-- two dimensional spaces with zero mean curvature. Numerous investigations on various aspects
of this special case have been carried out in the past \cite{ellisbefore1, ellisbefore2, ellisbefore3, ellisbefore4} and also
recently \cite{ellisrecent1, ellisrecent2, ellisrecent3, ellisrecent4, ellisrecent5,  ellisrecent6,  ellisrecent7, ellisrecent8,  ellisrecent9, ellisrecent10, ellisrecent11, ellisrecent12, ellisrecent13}. It has been suggested
that Lorentzian wormholes in general and BE geometry (including
its variants) in particular, could be useful
examples of black hole mimickers in the  context of gravitational
wave physics \cite{bhm1,bhm2,bhm3,bhm4,bhm5,bhm6,bhm7}.

\noindent The original BE spacetime, as evident from (1), 
does have a nontrivial
`redshift function' (i.e. $g_{00} = -e^{f(l)} $). Here we consider a different `redshift function' which ensures a local violation
(around the wormhole throat) of only one of the Null Energy Condition (NEC)
inequalities. We shall see that the Averaged Null Energy Condition (ANEC) violation may also be controlled
by choosing appropriate values of the metric parameters (eg. 
macroscopic wormholes with large throat radii). 
Further, as we show and discuss below, our
geometry has a link (via a non-vacuum generalisation and a static limit) with the Witten bubble spacetime
\cite{witten} known in five dimensional,  vacuum Kaluza--Klein theory.
Finally, with the chosen `red-shift function',
geodesics
and scalar waves turn out to be exactly solvable, thereby leading to 
tractable as well as interesting consequences.

\noindent To begin, we first state the line element. It is
given as,
\begin{eqnarray}
ds^2 = -\alpha^2 r^2 dt^2 + \frac{dr^2}{1-\frac{b_0^2}{r^2}} +  r^2 d\Omega_2^2
\end{eqnarray}
where $b_0$ and $\alpha$ are two distinct parameters.
Asymptotically, the above line element becomes that of a {\em non-flat, spherical Rindler} type
spacetime. 
Recall that the spherical Rindler spacetime
mentioned in
\cite{culetu, vijay} is globally flat and can be rewritten as 
Minkowski spacetime via a global coordinate transformation. However, the
spacetime obtained in the asymptotic $r\rightarrow \infty$ limit of the above-stated line element has a {\em Rindler $t-r$ section} 
(i.e. $ ds^2 = \left [ -\alpha^2 r^2 dt^2 + dr^2 \right ] + r^2 d\Omega_2^2$) but is not globally flat.
As we will see later, several curvature scalars vanish as 
$r\rightarrow \infty$ and are
large but finite near the throat even though the metric is not
asymptotically flat. This feature is not uncommon and has been noted
in various examples, most notably, say in the so--called Kiselev black holes \cite{kiselev, visser} or in the metrics 
discussed in \cite{clement}. Since $r\geq b_0$, there is no singularity anywhere.
The spatial section which is indeed asymptotically flat, has, visibly, the features of the $M=0$ BE wormhole. 

\noindent The specific form of $g_{00}$ in the line element in (5) is directly related to
the existence of a conformal Killing vector in the geometry. This has been shown 
and used in \cite{lobo, kuhfittig} in the context of wormholes, though the
original idea  appears in earlier papers \cite{ponce, roy}.  
Unfortunately, the fact that the line element is not asymptotically flat seems to have
been a deterrant in either using it in the wormhole context or analysing its
properties further.  Our purpose, in this article, is to fill this gap by 
studying some of the
interesting features of the geometry in (5), which seems to have escaped
attention.

\noindent One may physically understand the influence of the
chosen $g_{00} = -\alpha^2 r^2$ by calculating the frequency
shift of a light signal: (a) emitted from the throat and ending up at
larger values of $r$, or (b) emitted from a larger value of
$r$ and reaching the throat at $b_0$ later. For (b), one encounters
a {\em blue-shift} whereas in (a) there is a {\em red-shift}. Thus, 
an observer at a finite location/infinity would perceive the presence of a wormhole
throat via a finite/infinite red shift. This is in contrast  to
what happens for a black hole where the presence of a horizon 
is manifest through an infinite red-shift (the reason behind the name
`red-shift function') of a signal emitted from the horizon, as seen by {\em any} observer away from the horizon.

\noindent We now move on to further discuss different aspects
of this spacetime.

\section {\bf The spacetime geometry and matter}

\noindent In order to keep things somewhat general let us focus on the
following line element:  
\begin{eqnarray}
ds^2 = -\alpha^2 r^2 dt^2 + \frac{dr^2}{1-\frac{b(r)}{r}} + r^2 d\Omega_2^2
\end{eqnarray}
where $b(r)$ is kept unspecified, as of now. 
Evaluating the Einstein tensor (we do it in the frame basis) and defining $T_{ij} = \frac{1}{\kappa} G_{ij}$ 
(from General Relativity(GR)) one can arrive at the NEC inequalities (for a diagonal $T_{ij}$ with $T_{00} =\rho$,
$T_{11} =\tau$, $T_{22} = T_{33} = p$, NEC gives $\rho+\tau \geq 0, \rho+p \geq 0$):
\begin{eqnarray}
\rho +\tau \geq 0 \Rightarrow \frac{b'r- 3b}{r^3} + \frac{2}{r^2} \geq 0 \\
\rho+p \geq 0 \Rightarrow \frac{1}{r^2} \geq 0 
\end{eqnarray}
The second inequality is trivially true while the first one, as we will see for the specific (BE) choice
of $b(r) = \frac{b_0^2}{r}$, yields
\begin{eqnarray}
\rho+\tau \geq 0 \Rightarrow \frac{2r^2-4b_0^2}{r^4} \geq 0
\end{eqnarray}
Thus the $\rho+\tau \geq 0$ inequality will hold good as long as $r^2 \geq 2 b_0^2$, leading to
a finite violation in the region $b_0 \leq r < \sqrt{2} b_0$
\cite{lobo} (see also \cite{azreg} for discussion on localised NEC violations). It is clear that this 
localised violation happens only because of the red-shift function we have 
chosen in this variant of the standard $M=0$ BE line element (see Fig. 1). The
function $\rho+\tau$ has a zero at a location beyond $b_0$ and this happens because
of the choice of $g_{00}$. In principle, other choices are also possible, for
example $g_{00} = -\left (\alpha r \right )^m$ with $m\geq 2$ ($m=2$ being the case
considered here). 
\begin{figure}[h!]
\epsfxsize=4.5in\epsffile{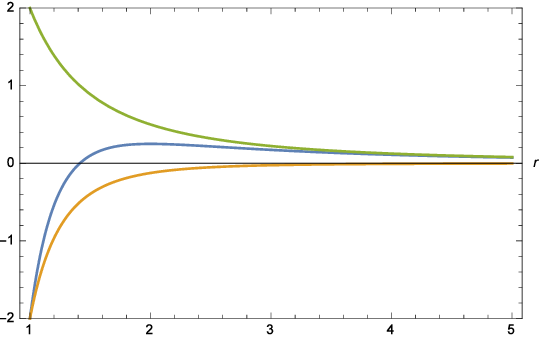} \caption{{NEC inequality $\rho+\tau$ ($y$ axis) versus $r$ ($x$ axis) for
$M=0$ BE (yellow), the variant of BE (blue) and the non-flat spherical Rindler type (green).
Note that for BE, violation is for all $r$, for the variant of BE it happens
around the throat and for the non-flat spherical Rindler type there is no violation.
}} \label{fig1}
\end{figure}

\noindent Let us now turn to evaluating the ANEC given by
the integral $\int_{\lambda_1}^{\lambda_2} T_{ij} k^i k^j d\lambda \geq 0$. A quick
calculation along radial null geodesics (see \cite{mv}) gives the value of the ANEC integral as:
\begin{eqnarray}
\frac{1}{\kappa \alpha} \int_{b_0}^{\infty} \frac{2r^2-4 b_0^2}{r^4 \sqrt{r^2-b_0^2}} \, dr = -\frac{2}{3 \kappa \alpha b_0^2}
\end{eqnarray}
Hence, the ANEC is also violated. However, as is evident from the 
expression, for a fixed $\alpha$, the value of the integral becomes smaller and
smaller for larger $b_0$ (i.e. for macroscopically large wormholes). 
This is expected and not really unusual. If $g_{00}=-1$ (i.e. a standard
$M=0$ BE wormhole) the ANEC integral would have a value equal to
$-\frac{\pi}{2 \kappa b_0}$. It is indeed clear that one may use both $\alpha$
and $b_0$
in order to tune the ANEC integral to very small (but negative) values. 

\noindent As in other scenarios studied earlier {\cite{bhm6}}, we may write the
matter required to support the geometry as a sum of two parts--one due to a
phantom scalar (violating the NEC) and another satisfying the NEC. More precisely, if we write
$\rho =\rho_\phi + \rho_e, \tau=\tau_\phi+\tau_e$ and $p=p_\phi+p_e$, we have
\begin{eqnarray}
\rho = \rho_\phi+ \rho_e = \frac{1}{\kappa} \left [ \left (-\frac{b_0^2}{r^4} \right ) + 0 \right ] \\
\tau=\tau_\phi +\tau_e = \frac{1}{\kappa} \left [ \left (-\frac{b_0^2}{r^4} \right ) + 
\left (\frac{2r^2-2b_0^2}{r^4}\right ) \right ] \\
p=p_\phi + p_e = \frac{1}{\kappa} \left [ \left ( \frac{b_0^2}{r^4} \right ) + \left (\frac{1}{r^2}\right )\right ]
\end{eqnarray}
where the first terms inside square brackets in the R.H.S. are due to the phantom scalar (which generates the
Bronnikov--Ellis geometry) and the rest define the extra piece which is NEC satisfying.
This split clearly shows how the redshift function plays a role in 
confining the NEC violation around the throat only.

\noindent The curvature properties of this geometry may be
noted through the behaviour of the Ricci ($R$) and Kretschmann ($K$) 
scalars given as,
\begin{eqnarray}
R = -\frac{4}{r^2} \hspace{0.2in};\hspace{0.2in} K = \frac{8}{b_0^4} 
x^4 \left [ x^2 (3 x^2-2) +1\right ] 
\end{eqnarray}
where $x=\frac{b_0}{r}$ and $0\leq x \leq 1$. We find that both $R$ and
$K$ tend to zero as $r\rightarrow \infty$. Moreover, the Ricci scalar $R$
is independent of the value of $b_0$ and is manifestly negative 
at all finite $r$. $K$, of course, is everywhere positive.

\noindent One may be tempted to analyse the NEC and ANEC for
a wider class of functions $b(r)$ labeled with
a parameter $\nu$ and given as:
\begin{eqnarray}
b(r) = b_0^\nu r^{1-\nu}
\end{eqnarray}
With such a choice,  the violation of the local NEC
around the throat persists in the region
$b_0 \leq r \leq \left (1+\frac{\nu}{2}\right )^{\frac{1}{\nu}} b_0$. The ANEC integral
after evaluation gives:
\begin{eqnarray}
\frac{1}{\kappa\alpha}\int_{b_0}^{\infty} \frac{\left (2 r -(\nu+2) b_0^\nu r^{1-\nu} \right ) r^{\frac{\nu}{2}}}{ r^4 \sqrt{r^\nu -b_0^{\nu}}} dr = - \frac{1}{\kappa\alpha b_0^2}
\frac{\sqrt{\pi} \Gamma \left (\frac{2}{\nu}\right )}{\nu \Gamma \left ( \frac{3}{2} +\frac{2}{\nu} \right )}
\end{eqnarray}
We can verify that as $\nu$ becomes larger in value
the negativity of the integral decreases (see Fig. 2). Also,
as observed earlier, for all $\nu$, the violation
for fixed $\alpha$ scales as $\frac{1}{b_0^2}$, i.e. for large wormholes the ANEC violation is less.
\begin{figure}[h!]
\epsfxsize=4.5in\epsffile{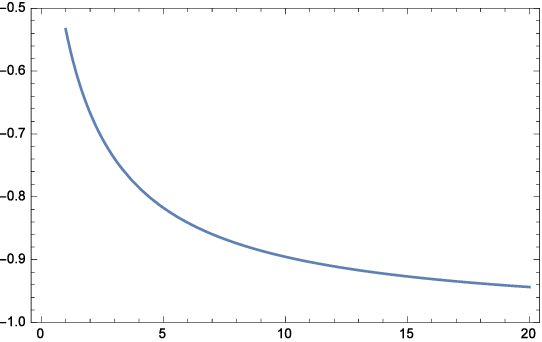} \caption{{ANEC integral without the
factor $\frac{1}{\kappa\alpha b_0^2}$ ($y$ axis) as a function of the metric parameter $\nu$ ($x$ axis).}} \label{fig1}
\end{figure}

\noindent Thus, in retrospect, we may say that we have
variant of the BE line element which represents a Lorentzian
wormhole (or a class of wormholes with a one-parameter family of $b(r)$)
with a non-trivial red-shift function.
Our spacetime requires matter with a
local (near the throat) violation of the NEC and an ANEC violation 
which can be controlled by adjusting the metric parameters (eg. 
macroscopically large throat radius). 

\noindent It is natural to ask--why do we need to
bother about such a line element? 
Is it just to ensure a localised NEC violation? 
We now elaborate on a precise reason (different from the 
conformal Killing vector analysis in \cite{lobo, kuhfittig}) behind 
this choice of the red-shift function (i.e. $g_{00}$),
thereby providing our motivations.

\section {\bf Non-vacuum Witten bubble--like extensions}

\noindent About four decades ago, Witten \cite{witten} found a
vacuum bubble spacetime in five dimensional Kaluza-Klein
theory by performing a double 
Wick rotation  \cite{witten}, \cite{bachelot} of 
five dimensional Schwarzschild geometry ( $T=i\chi$ and $\Theta = i t +\frac{\pi}{2}$, where $T, r, \Theta, \Phi, \xi$
are coordinates in the 5D Schwarzschild). The idea there
was to demonstrate an instability in the Kaluza--Klein
vacuum via this construction. Witten's spacetime is given as
\begin{eqnarray}
ds^2 = -r^2 dt^2 +\frac{dr^2}{1-\frac{b_0^2}{r^2}} + r^2 \cosh^2 t \, d\Omega_2^2 + \left (1-\frac{b_0^2}{r^2} \right ) d\chi^2
\end{eqnarray}
where $\chi=R \xi$ and $0\leq \xi \leq 2 \pi$ (periodic), $R$ a constant. Also $b_0\leq r < \infty$. The spacetime
topology is $R^2\times S^2\times S^1$.
Geodesics \cite{brill} and scalar waves
\cite{biplab} in the Witten bubble have been
studied in some detail in the past.

\noindent One can generalise this spacetime to a non-vacuum scenario by introducing new
parameters. Such a variant could be
given by a line element,
\begin{eqnarray}
ds^2 = -\alpha^2 r^2 dt^2 +\frac{dr^2}{1-\frac{b_0^2}{r^2}} + r^2 \cosh^2 \rho_1 t \, d\Omega_2^2 + \left (1-\frac{b_0^2}{r^2} \right ) d\chi^2
\end{eqnarray}
where $\rho_1 \neq \alpha >0$.

\noindent The purpose behind our construction is to
eventually take a limit $\rho_1 =0$, which will reduce
the spacetime in (18) to a static one. Such a static spacetime
will have its $\chi=$constant section as the
four dimensional geometry we have proposed in the earlier sections of
this article. Note that the Witten bubble--like spacetime
in (18)
cannot be obtained as a double Wick rotation
of 5D Schwarzschild, more so because it is
non-vacuum.

\noindent It will be useful to write down the Einstein
tensor for this generalised spacetime in (18) and note its
non--vacuum character as well as the nature of the
required matter {\em vis-a-vis} the energy conditions (NEC).
We find (in the frame basis and using five dimensional
GR),
\begin{eqnarray}
T_{00} = \frac{1}{\kappa_5} G_{00} =  \rho = \frac{1}{\kappa_5 r^2} \left (\frac{\rho_1^2}{\alpha^2} -1 \right ) \tanh^2 \rho_1 t \\
T_{11} = \frac{1}{\kappa_5} G_{11} = \tau =  - \frac{1}{\kappa_5r^2} \left (\frac{\rho_1^2}{\alpha^2} -1 \right ) \left (2+ \tanh^2 \rho_1 t \right ) \\
T_{22} = T_{33} = p=\frac{1}{\kappa_5} G_{22} = - \frac{1}{\kappa_5 r^2} \left (\frac{\rho_1^2}{\alpha^2} -1 \right ) \\ 
T_{44}  = p_e = \frac{1}{\kappa_5} G_{55} = - \frac{1}{\kappa_5 r^2} \left (\frac{\rho_1^2}{\alpha^2} -1 \right ) \left (2+ \tanh^2 \rho_1 t \right )
\end{eqnarray}
where $\kappa_5$ is the five dimensional generalisation of
$\kappa$. 

\noindent It is easy to verify that the NEC ($\rho+\tau \geq 0, \rho+ p \geq 0,
\rho+ p_e \geq 0$) holds good as long as $\rho_1 \leq \alpha$. 
For $\rho_1=\alpha$ we get back the
vacuum Witten bubble spacetime.
On the other hand, the generalised spacetime has a viable 
$\rho_1 =0$ limit (since $\rho_1$ and $\alpha$ are independent 
parameters). In this limit, we have a static spacetime given as:
\begin{eqnarray}
ds^2 = -\alpha^2 r^2 dt^2 +\frac{dr^2}{1-\frac{b_0^2}{r^2}} + r^2 d\Omega_2^2 + \left (1-\frac{b_0^2}{r^2} \right ) d\chi^2
\end{eqnarray}
for which the matter required has the following peculiar character:
\begin{eqnarray}
\rho =0 \hspace{0.2in} ; \hspace{0.2in} \tau = \frac{1}{\kappa_5} \frac{2}{r^2}\\
p= \frac{1}{\kappa_5} \frac{1}{r^2} \hspace{0.2in} ; \hspace{0.2in} 
p_e = \frac{1}{\kappa_5} \frac{2}{r^2}
\end{eqnarray}
Interestingly, this stress-energy also satisfies the NEC. Notice the complete
absence of $b_0$ in the expressions in (19)--(22) and also in (24)--(25). This happens only when we choose $b(r) = \frac{b_0^2}{r}$, not for all $b(r)$, as we can easily see
from the following discussion.

\noindent Let us try to understand how the NEC holds by considering a somewhat general
five dimensional spacetime given as:
\begin{eqnarray}
ds^2 = -\alpha^2 r^2 dt^2 +\frac{dr^2}{1-\frac{b(r)}{r}} + r^2 d\Omega_2^2 + \left (1-\frac{b(r)}{r}
\right ) d \chi^2
\end{eqnarray}
where $b(r)$ is an un-specified function to start with. If we write down the
NEC inequalities $\rho+\tau \geq 0$, $\rho+ p\geq 0$ (these are the only two which are relevant
since $\tau=p_e$) we can check that we obtain
\begin{eqnarray}
\rho+\tau \geq 0 \Rightarrow \frac{b''r^2-b' r- 3b}{2r^3} + \frac{2}{r^2} \geq 0\\
\rho+p \geq 0 \Rightarrow \frac{1}{r^2} \geq 0
\end{eqnarray}
In the usual four dimensional wormhole spacetime (say, the one considered earlier), the first of the above inequalities (Eqn. (27), i.e.
the $\rho+\tau \geq 0$ inequality) results in the
requirement $b'r-b\geq 0$, near the wormhole throat. 
This contradicts the 
requirement on a wormhole shape as found from embedding features, namely,
$b-b'r >0$ \cite{ellisbefore1}. In the five dimensional geometry
considered here, the $\rho+\tau \geq 0$, does not quite contradict
the embedding criterion for a wormhole shape, as mentioned above.  
For example, if we choose a class of functions given as
$b(r) = b_0^\nu r^{1-\nu}$ we find the inequality in (27) to be
\begin{eqnarray}
\frac{1}{2 r^2} \left ( \frac{b_0}{r} \right )^\nu \left (\nu^2-4\right ) +\frac{2}{r^2} \geq 0
\end{eqnarray}
which always holds for all $\nu\geq 2$! In principle, there could be many functions $b(r)$
for which the NEC will be satisfied. 

\noindent The above five dimensional static spacetime can indeed serve as
an example of a five dimensional wormhole satisfying the NEC (in fact, all
energy conditions). Earlier work \cite{culetu} addressed
the wormhole features of the original non--static Witten-bubble
spacetime. Here, we have introduced a parameter $\rho_1$, using which
we can obtain a static five dimensional wormhole. 
The nature of the matter could be questioned--the 
energy density is identically zero, as seen from the frame of a static observer!
Despite this peculiarity, the spacetime probably provides us with a counterexample
(albeit via extra dimensions and perhaps, entirely mathematical) where energy condition violating matter
does not seem to be a pre-condition for the existence of a wormhole
in GR with extra dimensions. Note that near the throat the extra dimension
vanishes whereas in the asymptotic regions it exists with a constant radius. A constant $\chi$ (constant extra dimensional) section 
of the five dimensional static geometry is the spacetime we 
have been talking about in our previous sections. 
The meaning and relevance of the higher dimensional spacetimes
mentioned in this section may surely
be explored further in future.

\section{\bf Geodesic motion} 

\noindent Let us now return to the original four dimensional
spacetime which appears as a $\chi=$constant section of the static, five dimensional spacetime mentioned above. We now try to find the timelike and null geodesics in this four dimensional spacetime.

\noindent The null and timelike geodesics satisfy the
condition,
\begin{eqnarray}
- \alpha^2 r^2 {\dot t}^2 + \frac{{\dot r}^2}{1-\frac{b_0^2}{r^2}} +
r^2 {\dot \theta}^2 + r^2 \sin^2 \theta {\dot \phi}^2 = -\gamma
\end{eqnarray}
where $\gamma=1$ for timelike and $\gamma=0$ for null. The overdot here
denotes a derivative w.r.t. the parameter $\tau$ labeling points on
the geodesics.

\noindent We may work with the choice $\theta= \frac{\pi}{2}$ 
since it satisfies the 
$\theta$ geodesic equation. We also have two constants of motion 
$E$ and $L$ given as:
\begin{eqnarray}
\dot t  = \frac{E}{\alpha^2 r^2} \hspace{0.2in} ; \hspace{0.2in} 
\dot \phi = \frac{L}{r^2}
\end{eqnarray}
Using the above in the timelike geodesic condition ($\gamma =1$) we
arrive at the following equations:
\begin{eqnarray}
\frac{dr}{dt} = \pm \frac{\alpha^2}{E} \sqrt{\left (d_0^2-r^2\right )\left (r^2-b_0^2\right )} \\
\frac{d\phi}{dt} = \frac{L}{E} \alpha^2
\end{eqnarray}
where $d_0^2 = \frac{E^2}{\alpha^2} - L^2 > 0 $.

\noindent One may interpret the expression for $\frac{dr}{dt}$ using an
effective potential. This leads to the definition of $V(r)$:
\begin{eqnarray}
\left (\frac{dr}{dt}\right )^2 = - \frac{\alpha^4}{E^2} \left (r^2 -d_0^2\right ) \left (r^2-b_0^2\right ) = -V(r)  
\end{eqnarray}
Fig. 3 (figure on the left) shows the graph of $V(r)$ where the $V=0$ 
horizontal line intersects $V(r)$
at $b_0$ and $d_0$ (endpoints of the periodic motion, see below). $V(r)$ here is a well
potential whereas in $M=0$ BE
geometry the corresponding $V(r)$ has a barrier near the
throat--the difference exclusively driven via the
chosen redshift function.

\noindent We first note that
circular orbits are possible for $r=d_0$ and
also at $r=b_0$ (with a constraint $b_0^2 \alpha^2 = E^2- \alpha^2 L^2$).  

\noindent  General periodic solutions to the $r$ equation exist for both $
L=0$ and $L \neq 0$. The solution ($L\neq 0$) is given as:
\begin{eqnarray}
r(t) = b_0 \, nd \left [ d_0 \frac{\alpha^2}{E} t ; \frac{d_0^2 -b_0^2}{d_0^2} \right ]  \\
\phi (t) = \frac{L \alpha^2}{E} t 
\end{eqnarray}
where `$nd[x,k]$' is a periodic Jacobian elliptic function with real periodicity
given by $2 K(k)$ ($K(k)$ denotes the elliptic integral of the third kind). Fig. 3 (figure on the right side)
shows a graph for $r(t)$. It is possible to integrate the relation $\dot t =\frac{E}{\alpha^2 r^2}$ and obtain an expression of $\tau (t)$. We have checked
(not presented here)
that $\tau(t)$ is monotonically increasing and given in terms
of various Jacobi elliptic functions. 
Note further that
in the limit $d_0\rightarrow b_0$ (i.e. $k\rightarrow 0$)
one gets back the circular orbit $r=b_0$ because
$nd \left [ x,0 \right ] =1$.

\begin{figure}[ht]
\begin{minipage}[b]{0.45\linewidth}
\centering
\includegraphics[width=\textwidth]{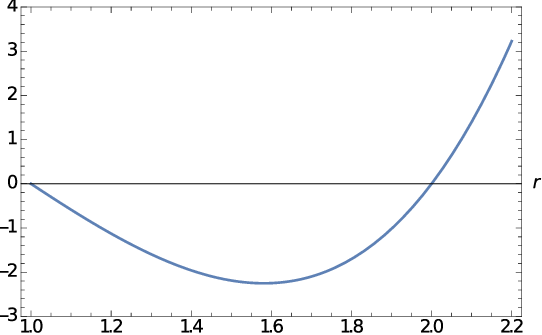}
\label{fig:figure1}
\end{minipage}
\begin{minipage}[b]{0.45\linewidth}
\centering
\includegraphics[width=\textwidth]{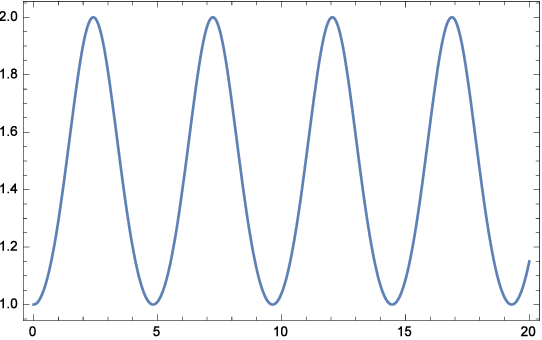}
\label{fig:figure2}
\end{minipage}
\caption{Left: Effective potential $V(r)\times E^2$. Right: $r(t)$ as in Eqn. (35).
Values for both figures: $\alpha=1$, $L=1$, $E=\sqrt{5}$, $b_0=1$, $d_0=2$. The right figure
represents the Jacobian elliptic function `nd'.}
\end{figure}


\noindent We may also write the solution as
\begin{eqnarray}
r(\phi) = b_0 \, nd \left [ d_0 \frac{\phi}{L} ; \frac{d_0^2 -b_0^2}{d_0^2} \right ]
\end{eqnarray}
For closed orbits, we require:
\begin{eqnarray}
\frac{2 \pi d_0}{L} = 2 K(k)
\end{eqnarray}
where $k=\frac{d_0^2-b_0^2}{d_0^2} < 1$.

\noindent In time $t_c = \frac{2 K(k) E}{d_0 \alpha^2}$ the particle returns to
$r=b_0$ after a full circuit in $\phi$. To
illustrate this feature on closed and open (attractor) orbits we give the following
example. Take $d_0=2$, $\alpha=1$, $E=\sqrt{13}$ and $L=3$. The condition for
closed orbits turns out to be: $K(k)=\frac{2 \pi}{3}$. Solving, we get $k= 0.712795$.
Hence, $b_0 = 1.07183$. Using these numerical
values one can obtain the blue curve in Fig 4 which represents a
closed orbit with minimum and maximum $r$ given as
$r=1.07183$ and $r=2$. The yellow curve in Fig. 4 is for a
different set of parameters.
In contrast, if we do not use the condition
$K(k) = \pi \frac{d_0}{L}$ in order to find $k$,
we end up with an open orbit for which the
values of $r$ are confined between $b_0$ and
$d_0$ (see Fig. 5 where $d_0=2$, $b_0=1$, $k=0.75$, $L=1$, $\alpha=1$ and $E=\sqrt{5}$,
as in Fig 3.). Thus, both closed and open
periodic orbits exist under specific 
conditions.
\begin{figure}[h!]
\epsfxsize=3.5in\epsffile{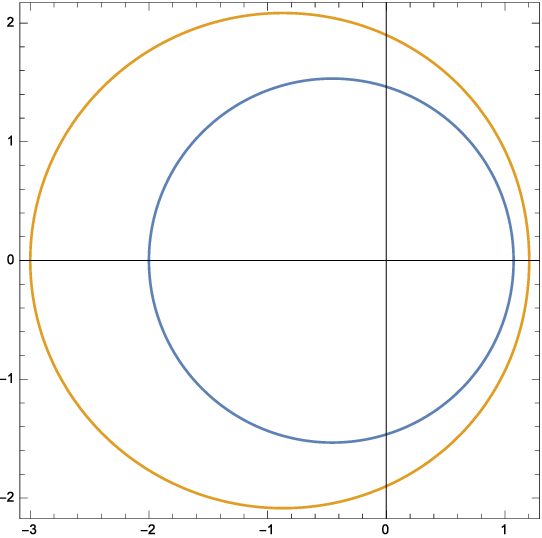} \caption{
Polar plot of closed orbits with (i) $d_0=2$, $b_0=1.07183$, $L=3$, $E=\sqrt{13}$, $\alpha=1$, $k=0.712795$ [blue], (ii) $d_0=3$, $b_0=1.20342$, $L=4$, $E=5$, $\alpha=1$ [yellow]. } \label{fig1}
\end{figure}
\begin{figure}[h!]
\epsfxsize=3.5in\epsffile{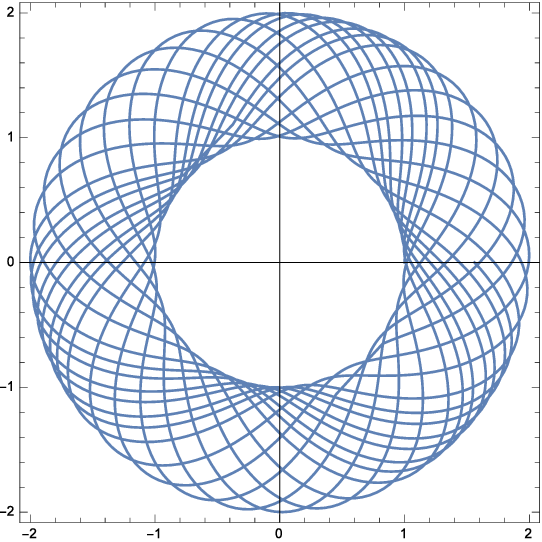} \caption{{ Polar plot of open, bounded periodic orbit for $k=0.75$, $L=1$, $E=\sqrt{5}$, $d_0=2$, $b_0=1$, $\alpha=1$.}} \label{fig1}
\end{figure}

\noindent When $L=0$, it is clear that $\phi$ is a constant. We also have
$d_0 = \frac{E}{\alpha}$. Thus, we end up with
\begin{eqnarray}
r(t) =  b_0 \, nd \left [ \alpha t ; \frac{E^2 - b_0^2 \alpha^2}{E^2} \right ]
\end{eqnarray}
The trajectory now is at a fixed $\phi$ and the particle oscillates
up and down from $b_0$ to $d_0$ and vice-versa. It is easy to see that
the particle starting from $r=b_0$ at $t=0$ will return to $r=b_0$ in time
$\alpha t_c  = 2 K(k)$. 

\noindent Thus, for both $L=0$ and $L\neq 0$ one has periodic motion
and we are able to find the functional dependence exactly. The closed orbits as well as the open bounded orbits found here are unique to
this generalised BE spacetime and is absent in
the standard (i.e. $f=0$) BE spacetime usually studied. 
Furthermore, it is
useful to note that the static geometry representing the
generalisation of the Witten bubble will have the same geodesics
for $\chi=$ constant (which solves the $\chi$ geodesic equation).

\noindent Another important and interesting quantity which can be
easily calculated is the expansion $\Theta$ of a timelike geodesic congruence generated 
by the  geodesic vector field $u^i$ given as:
\begin{eqnarray}
u^i \equiv \left  (\frac{E}{\alpha^2 r^2}, \frac{1}{r^2} \sqrt{(r^2-b_0^2)(d_0^2-r^2)}, 0, \frac{L}{r^2} \right )
\end{eqnarray}
A straightforward calculation leads to the expression for $\Theta$ as:
\begin{eqnarray}
\Theta = \nabla_i u^i = \frac{2b_0^2 \left (r^2-d_0^2\right ) - d_0^2 \left (r^2-b_0^2 \right ) + 3 \left (r^2-b_0^2\right ) \left ( d_0^2-r^2\right )}{r^3 \sqrt{\left(r^2 -b_0^2\right ) \left(d_0^2-r^2\right )}}
\end{eqnarray}
It is easy to see that in the $r\rightarrow b_0$ and $r\rightarrow d_0$ limits the
expansion $\Theta$ diverges to negative infinity, indicating benign focusing. 
In particular, a timelike geodesic congruence which starts out at
some point between $b_0$  and $d_0$ with an initially negative expansion will eventually
focus at $b_0$ or $d_0$--a result which follows from the well--known focusing theorem
obtained as a conclusion from the Raychaudhuri equation.

\noindent It is even more easy to find the null geodesics though it
is not periodic motion.
One can easily check that the radial null geodesics ($L=0$, $\phi$
constant) in this
spacetime are given as,
\begin{eqnarray}
r = b_0 \cosh \,\alpha t
\end{eqnarray}
The trajectory represents a photon starting out at the throat and
escaping to infinity, along a curve with $\phi$ fixed.

\noindent When $L\neq 0$, i.e. for the non-radial null geodesics, the
solution turns out to be,
\begin{eqnarray}
r(t) = b_0 \cosh \,\, \frac{ d_0 \alpha^2}{E} t \hspace{0.2in};\hspace{0.2in} \phi = \frac{\alpha^2 L}{E} t
\end{eqnarray}
Here, the photon spirals away to infinity because of the time variation of
$\phi$.

\noindent It is worth noting that the null geodesics 
found in this geometry are also different from those of
the BE geometry. 
For example, for $L=0$, in BE spacetime,
one gets 
\begin{eqnarray}
r(t) = r(\tau) = \sqrt{b_0^2 + t^2} = \sqrt{b_0^2 + E^2 \tau^2}.
\end{eqnarray}
which is functionally different from the expression in (42) above.

\noindent The periodic and non-periodic behaviour of timelike and
null geodesics respectively, is somewhat reminiscent of geodesics in
Anti-de Sitter (AdS) spacetime where the radial null geodesics
exhibit a runaway to infinity at finite $t$, whereas, the
timelike geodesics have periodic behaviour. A sort of 
reason behind this `similar'
behaviour could be attributed to the fact that $g_{00} = -\alpha^2 r^2 $ for our
spacetime and for AdS, it is simply $g_{00} = - (1+\alpha^2 r^2$).

\noindent Thus, one may say that all geodesics in the proposed spacetime wormhole are known exactly, which, we feel is a
satisfying feature for this geometry. We also obtain
a fairly wide variety of trajectories -- periodic, non-periodic,
open, bounded and closed. If at all, such geometries are
ever deemed to be observationally relevant, then the diverse behaviour of 
test particles, as described above, could yield useful signatures.

\noindent We now turn to the final topic of discussion--scalar wave propagation--
which, as we will show, is also exactly solvable!

\section{\bf Scalar waves} 

\noindent The propagation of massless scalar waves 
in this spacetime geometry is governed by the equation:
\begin{eqnarray}
\Box \phi =0
\end{eqnarray}
Studying such scalar waves amounts to an analysis of scalar perturbations.
One can consider the perturbations as those for the scalar which generates part of the 
matter required to support the geometry (see earlier discussion in Section II).

\noindent Assuming $\phi = T(t) R(r) Y (\theta, \phi)$, one can easily separate variables
and look for solutions with $T(t) = e^{\pm i \omega t}$. 
The angular part $Y(\theta,\phi)$ is given in terms of the 
spherical harmonics $Y_{mn} (\theta,\phi)$. Using new coordinates
$r= b_0 \cosh \xi $, one arrives at the equation for $R(\xi)$ as
\begin{eqnarray}
\frac{d^2 R}{d\xi^2} +  2 \tanh \xi \frac{dR}{d\xi} + \left ( m(m+1) + \frac{\omega^2}{\alpha^2} \right ) R =0
\end{eqnarray}
A further substitution $ R = sech \xi  \, A (\xi)$ results in a surprisingly
simple equation for $A$:
\begin{eqnarray}
\frac{d^2 A}{d\xi^2} + \left (\frac{\omega^2}{\alpha^2} + m (m+1) -1\right ) A
=0
\end{eqnarray}
The final solution for $R(r)$ is therefore given as:
\begin{eqnarray}
R(r) = \frac{b_0}{r} \left ( C_1 \sin \left [ p \, {\cosh}^{-1} \frac{r}{b_0} \right ]
+ C_2 \cos \left [ p\, {\cosh}^{-1} \frac{r}{b_0} \right ] \right )
\end{eqnarray}
where,
\begin{eqnarray}
p = \sqrt{\frac{\omega^2 -\alpha^2}{\alpha^2} + m(m+1)}
\end{eqnarray}
and $C_1$, $C_2$ are integration constants.
Note that for $m=0$, $p$ is real (imaginary) for $\omega > \alpha$ ($\omega<\alpha$).  For all other $m\neq 0$, $p$ is real.
\begin{figure}[h!]
\epsfxsize=4.5in\epsffile{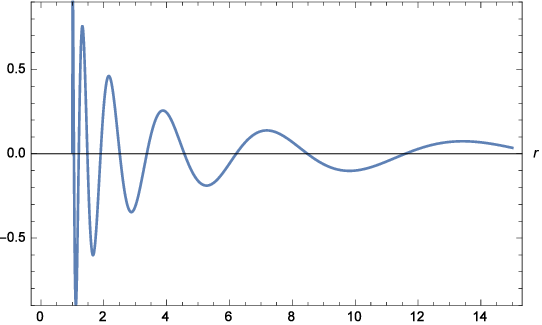} \caption{The radial wave function $R(r)$, assuming $\alpha=1$, $b_0=1$, $\omega^2 = 99 $, $C_2=0$, $C_1=1$} \label{fig1}
\end{figure}
\noindent The solution represents spherical waves with decreasing amplitude
as one moves away from the throat towards infinity (see Fig. 6). 
Thus, the spacetime is stable against such scalar perturbations
-- a fact which adds to its viability as a geometry worth considering.
Surely, more general perturbations (eg. axial and polar gravitational 
ones) have to be explored in order to understand the stability issue in a
more complete way. 

\noindent We may also contrast the behaviour of scalar wave propagation
in this geometry with corresponding scenarios in the asymptotic
nonflat, spherical Rindler type spacetime
and in the $M=0$ BE spacetime. In the non-flat spherical Rindler case, the
radial part $R(r)$ of $\phi$ is generically given by power laws of the
type $r^{-s}$ ($s>0$) -- thus the decay is not oscillatory. In the Bronnikov-Ellis spacetime, the
solution of the equation for $R(r)$ is given by the rather complicated 
radial oblate spheroidal functions, as noted and discussed many years ago 
in \cite{ellisbefore2}.

\section{\bf Concluding remarks}

\noindent Among many possibilities that emerge from this
work, one which could be quite important concerns
the geometry arising as the static limit of the
generalised Witten bubble--like spacetime in 5D, which can be
supported with matter satisfying the NEC. This
spacetime has wormhole features and 
its constant extra dimension section is the
4D geometry discussed here. 
We have shown that the presence of the
extra dimension as well as the 
choice of the red-shift function, somehow manages to evade the
NEC violation in this higher dimensional spacetime. It is clear that this feature will
persist with more extra dimensions. 
In addition, a further generalisation of Eqn. (23)
with the higher dimensional line element rewritten as,
\begin{eqnarray}
ds^2 = -e^{2 \psi(r)} dt^2 +\frac{dr^2}{1-\frac{b(r)}{r}} + r^2 d\Omega_2^2 + \left (1-\frac{d(r)}{r}
\right ) d \chi^2
\end{eqnarray}
where $b(r)\neq d(r)$ and $\psi(r)$ an additional function,
also seems to suggest viable wormhole spacetimes (for chosen $b(r)$, $d(r)$ and $\psi(r)$) in vacuum or with any required matter satisfying/violating the NEC \cite{sk2021}.
Thus, our work indicates a fairly
broad class of examples of NEC satisfying wormholes within the framework of higher dimensional GR. Strangely, if one considers the four dimensional
section (constant $\chi$) as an independent geometry (as we have done in this
paper) in four dimensional
GR, one ends up with localised NEC violation around the wormhole-throat. 
Further 
studies may help us in understanding
the above--mentioned intriguing features, in future.

\noindent It also turns out that geodesic motion as well as scalar wave propagation can be solved exactly for this spacetime, in terms of
known functions. This feature does not really exist in too many spacetimes
which arise in GR or other theories of gravity. 
In addition, the closed orbits and attractors
noted in timelike geodesic motion provides
useful signatures which may help in characterising the geometry.

\noindent Finally, a science fiction enthusiast may wonder if our 4D
spacetime wormhole is traversable. We have found that (not shown here), choosing the
dimensionless number $\alpha b_0$ appropriately we can indeed have a
traversable (for humans) wormhole. The tidal force constraints will, as
always, restrict the possible allowed value of the throat radius. However, it is necessary to understand the sense
of traversability here, with special reference to the fact that
timelike geodesics are all bounded between $r=b_0$ and $r=d_0$, though
the value of $d_0$ can be large via choices of $E$, $\alpha$ and $L$. Alternatively, one may consider \cite{lobo} taking a
timelike piece of this geometry, say, between $r=b_0$ and 
some $r= r_0> b_0$ and join the boundary at $r_0$ with
vacuum Schwarzschild. In such a case, one obtains an asymptotically flat traversable wormhole via cut-and-paste 
with a thin shell at $r_0$ 
having matter violating the energy conditions.

\noindent It goes without saying that our work is largely theoretical
(bordering on the exotic, to say the least!) and has no link with observations (as of now). However, we do feel that our
results provide an interesting outing, illustrating the links between
wormholes, energy conditions, extra dimensions on one hand
and exactly solvable geodesics and scalar wave propagation
on the other, with curious consequences which may perhaps motivate
future research. The role of the specially chosen 
`red-shift function' ($g_{00} = -\alpha^2 r^2$) and the
link with the Witten--bubble are both crucial and novel elements which have
guided our
pursuits in this article.

\end{document}